\documentclass[conference]{IEEEtran}
\usepackage{bm}
\usepackage{setspace}
\usepackage{cite}
\usepackage{amsthm}
\usepackage{epsfig}
\usepackage{amsmath}
\usepackage{amssymb}
\usepackage{latexsym}
\usepackage{enumitem}
\usepackage{mathtools}
\usepackage{amsfonts}
\usepackage{color}
\usepackage{optidef}
\usepackage{lettrine}
\usepackage{graphicx}
\usepackage{multicol,flushend,color}
\usepackage{wasysym}
\usepackage{float}
\usepackage{algorithm}
\usepackage[noend]{algpseudocode}
\usepackage{xcolor}
\usepackage{cuted}
\usepackage{url}
\usepackage[font=footnotesize]{caption}
\usepackage[font=footnotesize]{subcaption}

\newcommand{\E}{\ensuremath{\mathbb E}}

\begin{document}
	\title{Secure Deep-JSCC
	Against Multiple Eavesdroppers}
	
	\author{%
		\IEEEauthorblockN{%
			Seyyed 
			Amirhossein Ameli Kalkhoran$^\ast$\IEEEauthorrefmark{2}, 
			Mehdi Letafati$^\ast$\IEEEauthorrefmark{3},
			Ecenaz Erdemir\IEEEauthorrefmark{4}, 
			Babak Hossein Khalaj\IEEEauthorrefmark{3}, \\
		    Hamid Behroozi\IEEEauthorrefmark{3},  
			 and 
		Deniz G\"{u}nd\"{u}z\IEEEauthorrefmark{4}%
		}%
\\
		\IEEEauthorblockA{\IEEEauthorrefmark{3}  \small Electrical Engineering Department, Sharif University of Technology, Tehran, Iran}%
		\IEEEauthorblockA{\IEEEauthorrefmark{2} \small Computer Engineering Department, Sharif University of Technology, Tehran, Iran}%
		\IEEEauthorblockA{\IEEEauthorrefmark{4} \small Department of Electrical and Electronic Engineering, Imperial College London, UK}%
\IEEEauthorblockA{\small Emails:‌ $\ddagger$$\{$mletafati@ee., 
			behroozi@, 	khalaj@$\}$sharif.edu;  $\dagger$ameli@ce.sharif.edu; $\S$\{e.erdemir17, d.gunduz\}@imperial.ac.uk}
}
	
	\maketitle
	\def\thefootnote{*}\footnotetext{Equal contribution}\def\thefootnote{\arabic{footnote}}

	\IEEEaftertitletext{\vspace{-2.5\baselineskip}}

	\begin{abstract}  
In this paper,  a generalization of deep learning-aided joint source channel coding  (Deep-JSCC) approach to secure communications is studied.   
	We propose an end-to-end (E2E)  learning-based approach for 
		secure 
		communication  against
		\emph{multiple 
			eavesdroppers}
		over complex-valued fading channels.  
		Both scenarios of {colluding} and {non-colluding} eavesdroppers are studied.  
		For the colluding  strategy, eavesdroppers share their logits  to collaboratively infer private attributes  based on \emph{ensemble learning} method, while for   the  {non-colluding} setup  they act alone.  
		The goal 
		is to  prevent   
		eavesdroppers
		from inferring  private (sensitive) information about the transmitted images, while delivering  the images to a legitimate receiver with minimum distortion.  
		By generalizing the ideas of \emph{privacy funnel} and wiretap channel coding, the   trade-off between the image recovery  at the  legitimate node   
		and the information leakage to the eavesdroppers is characterized.  
		To solve  this \emph{secrecy funnel} framework,
		we implement deep neural networks (DNNs)  
		to 
		realize a data-driven secure communication scheme, without relying on a specific data distribution.  
		Simulations over CIFAR-10  dataset  verifies the secrecy-utility trade-off.  Adversarial accuracy of eavesdroppers are also studied over Rayleigh fading, Nakagami-$m$, and  AWGN channels to verify the \emph{generalization} of the proposed scheme.  
Our experiments show  that  employing the proposed secure neural encoding can decrease the adversarial accuracy by $28\%$. 
	\end{abstract}

	
	\IEEEpeerreviewmaketitle
	\vspace{-2mm}
	
	\section{Introduction} 
	\vspace{-1mm} 
	Driven by the growing interest in semantic communication systems \cite{SC},  intelligent   transmission of multimedia content
	has received much attention because of its various applications in  augmented/virtual reality (AR/VR),   Metaverse \cite{AdHoc}, and surveillance systems \cite{Im-retriev-deniz, caching}.      
	The adoption and success of such services rely highly on the security of the delivered contents---communication  systems  should understand the desired ``level of security'' and intelligently adapt the transmission scheme  accordingly \cite{arxive, medical}. 
	
Connected intelligence is foreseen  as the most significant driving force in the sixth generation (6G) of wireless communications.  To this end, artificial intelligence and machine learning  (AI/ML) algorithms are envisioned to be widely  incorporated  into  6G  networks, realizing an ``AI-native'' air interface.  
Nevertheless, security issues  at the \emph{wireless edge} of 6G networks are still  identified  as  open challenges \cite{6G-PLS}. 	
	The air interface of  6G  systems encounters   ever-rising   attacks, such as eavesdropping,  spoofing  \cite{WSKG-letter},  and man-in-the-middle  \cite{WSKG-GC}.

	Recently, a considerable  number of research has been dedicated to  the utilization of deep learning (DL) techniques 
		to optimize the performance of  wireless  systems,
	thanks to their outstanding performance and generalization capabilities  \cite{DJSCC-Deniz,BW-agile,vtc2022}.  
	In the context of  wireless security,  
 autoencoders (composed of linear layers) are exploited in  \cite{Eduard-AE}  
 over the additive white Gaussian noise (AWGN) wiretap channel.  
	To tackle the trade-off between the data rate and security, 
	a  weighted sum of block error rate and information leakage is used as the loss function (LF) for neural wiretap code design.   
	The data fed into the autoencoder is combined with additional  non-informative random bits 
	to 
	confuse the eavesdropper; while, this  also   reduces the communication  
	rate.   
Notably, most of the previous works, i.e., \cite{MI-estim-channel_coding,MI-Wiretap-estim, Eduard-AE},   focus on  learning-aided secure channel coding, rather than  
	taking into account the end-to-end (E2E) performance of  secure communications.  
	 The content of the transmitted data is not addressed in these works and the entire bit-stream  is equally treated as the secret information to be protected against an eavesdropper.  
	
The E2E communication of images from a source node  to a legitimate destination  can be considered as  a joint source channel coding (JSCC) problem. DL-aided JSCC design, a.k.a  Deep-JSCC, has received significant attention thanks to its  superior performance, particularly its lack of reliance on accurate channel state information \cite{DJSCC-Deniz}.  
	However, in JSCC, different from  separate   source and channel coding, the channel codeword is correlated  with the underlying source signal.  This can create vulnerabilities  in terms of leakage to eavesdroppers, despite  providing robustness against channel noise. 
	Inspired by \cite{AE-Deniz} and \cite{Ecenaz-icassp}, we provide a generalization of the Deep-JSCC approach   to secure communication problems against multiple eavesdroppers.
		In this regard, \cite{AE-Deniz} proposes  a generative adversarial network (GAN)-inspired secure neural encoder-decoder pair  over an AWGN wiretap  channel  against one eavesdropper.   The authors in \cite{Ecenaz-icassp}  propose  a variational autoencoder (VAE)-based approach for Deep-JSSC design  over binary symmetric channels, again considering  a single eavesdropper.  
		
\begin{figure}
		\vspace{0mm}
		\centering
		\includegraphics
		[width=3.75in,height=1.15in,trim={0.25in 0.2in 0.1in  0.35in},clip]{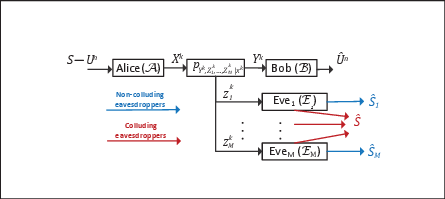}
		\vspace{-5mm}\caption{Proposed system model for secure  transmission against 
			multiple eavesdroppers.}
		\label{fig:SysModel}
  \vspace{-5mm}
	\end{figure}

		In this paper, we consider  \emph{E2E learning-based   secure communication  against  multiple eavesdroppers} for both colluding and non-colluding eavesdroppers,  over AWGN as well as  fading channels.  
For the  scenario of colluding  eavesdroppers, the adversaries share their logits to collaboratively infer private attributes based on  the \textit{ensemble learning} approach, while for   the  {non-colluding} setup  they act alone.   
	Please see Fig. \ref{fig:SysModel} for an illustration of the communication scenario studied in this paper.  
	Applications of our proposed framework include  (but are not limited to) 
	digital healthcare services, in which medical images   are sent to or received from  access points, while some private attributes, e.g., the identity of  patients should be kept  secret from potential eavesdroppers.  
	Notably, previous works \cite{AE-Deniz} and \cite{Ecenaz-icassp}  only considered a single eavesdropper with a single-antenna transmitter and multiple parallel channels, respectively. 
	In addition, both  \cite{AE-Deniz} and \cite{Ecenaz-icassp} are limited to static channels.  
	Moreover,  different from \cite{Eduard-AE}, no additional 
	redundant bits are required to be added to the source image.  
{In this paper, Rayleigh fading channel model is used  to represent time-varying  channel realizations during training, while  inference is performed over Nakagami-$m$ and AWGN channels, 
in addition to Rayleigh fading. 
Note that the encoder and the decoder  do not require any knowledge of the instantaneous channel gains in the proposed scheme.}	\footnote{\textit{Notations:} We denote the transpose, the conjugate transpose,  and $\ell^2$ norm of a vector by $(\cdot)^\mathsf{T}$, $(\cdot)^\dagger$,  and $||\cdot||$, respectively. 
The expected value and the probability density function (pdf) 
of random variable (RV) ${X}$ are denoted by $\E[{X}]$ and $p_X(x)$, 
respectively.  
Realization vectors of RVs are represented by bold lowercase letters. 
The mutual information of RVs ${X}$ and ${Y}$ and the cross-entropy  of two distributions $p$ and $q$ are shown, respectively, by $I({X};{Y})$ and $H(p,q)$.  
}


	\vspace{0mm}
	\section{System Model and Problem Statement} \label{sec:System_Model}	 
    \vspace{-1mm}
	Consider the  communication scenario  depicted  in Fig. \ref{fig:SysModel}, where a multi-antenna source  node, Alice ({$\cal A$}) with $n_{\sf T}$ antennas, aims to  deliver an image {${U}^n \in \mathcal{U}^n$}  to a  destination  node, Bob ({$\cal B$}), 
	over $k$  uses of 
	the  communication  channel, 
  {where $\mathcal{U}$ denotes the alphabet of source images.}     
	According to the JSCC literature \cite{DJSCC-Deniz}, we refer to  the image dimension, $n$,  as the  {source bandwidth}.  The channel dimension $k$  characterizes  the {channel bandwidth}, where we  usually have  $k < n$ to reflect the concept of {bandwidth compression}  \cite{DJSCC-Deniz}.   
	Image delivery should be kept secret from multiple eavesdroppers, denoted by Eve$_{1}$($\mathcal{E}_1$), $\cdots$,  Eve$_{M}$($\mathcal{E}_M$), which overhear the communication through 
	their own channels,  
	and want  to infer a private (sensitive)  attribute, e.g., diagnostic information regarding the source image, denoted by  $S \in \mathcal{S}$ with a discrete alphabet $\mathcal{S}$. 
{The eavesdroppers in the non-colluding setup act alone to infer the secret $S$,  
	while for the colluding setup,  ``knowledge sharing'' is also performed---they share  their logits based on    the concept of \emph{ensemble learning} 
	\cite{ensemble}.}

	\begin{figure*}
		\vspace{0mm}
		\centering
		\includegraphics
		[width=5.5in,height=2in,trim={0.1in 0.1in 0.1in 0.4in},clip]{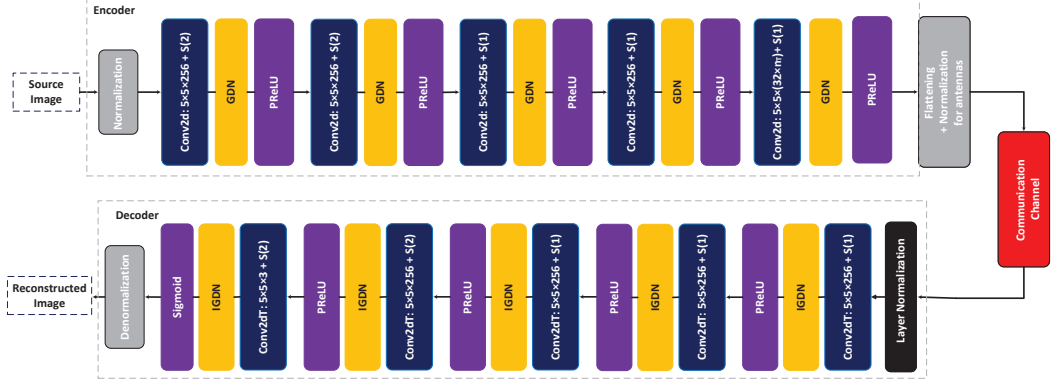}
		\vspace{0mm}
		\caption{Proposed DNNs at Alice (encoder) and Bob (decoder). 
			The notation $w\times w \times f$  denotes a convolutional layer with $f$ filters of spatial extent 
			$w$. 
			Moreover,  $s(\cdot)$ denotes  the stride, which  can be downsampling (at the encoder) or upsampling (at the decoder) \cite{DJSCC-Deniz}. At the output of the last 
			PReLU layer, which consists  of $2k \times n_{\sf T}$ elements, 
			we employ a flattening layer for each of the $n_{\sf T}$ antennas,  to reshape the  encoded tensor  to a data-stream.   The encoded latent sequence is further  normalized, such that the channel input satisfies the average transmit power constraint. }
		\vspace{-5mm}
		\label{fig:DNN}
	\end{figure*}

 
	Alice  maps the source information 
	$U^n$ into a {channel input codeword $X^k \in \mathbb{C}^{n_{\sf T} \times k}$} via an encoding 
	function {$f_{\cal A}: \mathcal{U}^n \rightarrow \mathcal{X}^k$}, where  $X^k = f_{\cal A}(U^n)$.  
	Transmitted codeword is subject to an average
	power constraint, $\frac{1}{k} \mathbb{E}[({X^k})^\dagger{X^k}] \leq P$, which will be satisfied for the realizations of the channel codeword in our data-driven approach.
	Channel outputs at $\mathcal{B}$ and $\mathcal{E}_m$  are denoted, respectively, by $Y^k$ and $Z_m^k \in \mathbb{C}^k$, with $m \in \{1, \ldots, M\}$.  
	Transmission of data-streams over the air  experiences  independent realizations of the  conditional   channel distribution $p_{Y, Z_1, \ldots, Z_M |‌X  }$.    
	We consider both the AWGN and  slow fading channels, where for the slow fading, we adopt two widely-used models of Rayleigh fading and Nakagami-$m$ channels, and {assume the channel realization to remain constant for the duration of the transmission of a single image, i.e., for $k$ channel uses.}
	Bob then applies a  decoding function
	$f_{\cal B}$ to obtain $\hat{U}^n = f_{\cal B}(Y^k)$. 
	Meanwhile, each Eve tries to extract the sensitive attribute  $S$, from her observations $Z_m^k$, or by collaborating with other Eves, i.e., sharing their logits. 
	We consider the trade-off between  delivering images
	to Bob with the highest fidelity and  controlling  the information leakage to each adversary, which  
	is theoretically 
	measured	by the mutual information metric  $I({S};{ Z}_m^k)$. 
Inspired by the key idea of \emph{privacy funnel} \cite{funnel},  we first formulate  an optimization   framework to characterize  this trade-off,  which we call \emph{secrecy funnel}. 
	This funnel-like framework is then  solved in a data-driven manner by implementing   DNNs.  
	Mathematically, we aim to simultaneously minimize both the distortion $d(U^n,\hat{U}^n)$ at Bob and the information leakage,  $I({S};{ Z}_m^k), m \in [M]$ to adversaries. 
	This corresponds to the following problem 
	\begin{align}\label{eq:P0}
		\underset{
			f_{\cal A},
			f_{\cal B}}{\text{minimize}}
		\quad 
		\mathbb{E}\left[d(U^n,\hat{U}^n)\right] + 
		\frac{1}{M}	\hspace{0mm}  \sum_{m \in [M]} w_m I({S};{ Z}_m^k),   
	\end{align}
where $w_m$  shapes our secrecy funnel and   adjusts the trade-off between  the information leakage and the distortion.     
Numerical solution of optimization \eqref{eq:P0}   is intractable due to the mutual information term.  Hence, we apply a variational approximation of mutual information, which was proposed in \cite{MI-approx}. Then, our optimization can be rewritten as:‌ 
	\begin{align}\label{eq:P1}
		\underset{f_{\Omega_{\cal A}},
			f_{\Omega_{\cal B}}}{\text{minimize}} \quad  \hspace{-2.5mm}\mathbb{E}\hspace{-0.5mm}\left[\hspace{-0.3mm}d(U^n,\hat{U}^n)\hspace{-0.3mm}\right] \hspace{-1mm} + \hspace{-1mm}‌
		\frac{1}{M}	\hspace{-2.5mm} \sum_{m \in [M]}  \hspace{-3mm}
		w_m
		\underset{q_{S|Z_m^k}}{\text{max}}\hspace{0mm}
		\hspace{-1mm}\mathbb{E}\hspace{-0.5mm}\left[\log q_{S|Z^{k}_{m}}(s|{\boldsymbol z}_m)\hspace{-0.5mm}\right],  
	\end{align} 
	where $q_{S|Z^k_m}(s|{\boldsymbol z}_m)$ characterizes the  
	Eve$_{m}$'s posterior  estimation corresponding to the correct
	distribution of $S$, 
	given the observation $Z^k_m = {\boldsymbol z}_m$. 	
To realize a data-driven approach,  $f_{\cal A}$ and $f_{\cal B}$ are parameterized by DNNs with parameters $\Omega_{\cal A}$ and $\Omega_{\cal B}$, respectively.  
Details of the DNN architectures and the  training strategies  are given in the next section.     
	The approximation  in \eqref{eq:P1} can be interpreted as the sample-wise negative  cross-entropy (CE)  between the distribution over  adversaries' predictions and the true 
	distribution of sensitive attributes. 
	To solve \eqref{eq:P1} in a data-driven manner, we assume that the 
	Eves also employ  adversarially-trained DNNs and try  to infer the sensitive attributes of the transmitted images, where  
	$\Theta_{E,m}$  parameterizes  the adversarial network of Eve$_{m}$. 
	Accordingly, we can formulate the following  LF.   
	\begin{align}\label{eq:P-DL}
		\hspace{-2mm} \mathcal{L}(\Omega_{\cal A},\Omega_{\cal B},&\mathbf{\Theta}_E)     = 
	   \mathbb{E}\left[d(U^n,f_{\Omega_{\cal B}}(Y^k))\right]  \nonumber
		 \\ 
		& + 
		\frac{1}{M}	\hspace{0mm} \sum_{i \in [M]} w_m \hspace{2mm} \underset{\Theta_{E,m}}{\text{max}}
		\hspace{1mm}
		\mathbb{E}\left[\log q_{\Theta_{E,m}}({s|{\boldsymbol z}_m})\right],  
	\end{align}
	where  $\mathbf{\Theta}_E \overset{\Delta}{=} (\Theta_{E,1}, \cdots,\Theta_{E,M})$, 
	and $q_{\Theta_{E,m}}({s|{\boldsymbol z}_m})$ formulates the approximated adversarial   likelihood regarding  the correct value $S=s$, 
	estimated  by the DNN of the  $m$'th adversary. 
	Details of the 
	proposed DNN architectures and the training strategies  are given in the following  section. 


	\vspace{-0mm}
	\section{DNN Architectures}\label{sec:Proposed_approach}
\begin{figure}
		\vspace{-3mm}
		\centering
		\includegraphics
		[width=2.85in,height=1.65in,
		trim={0.1in 1.75in 0.1in 1.0in},clip]{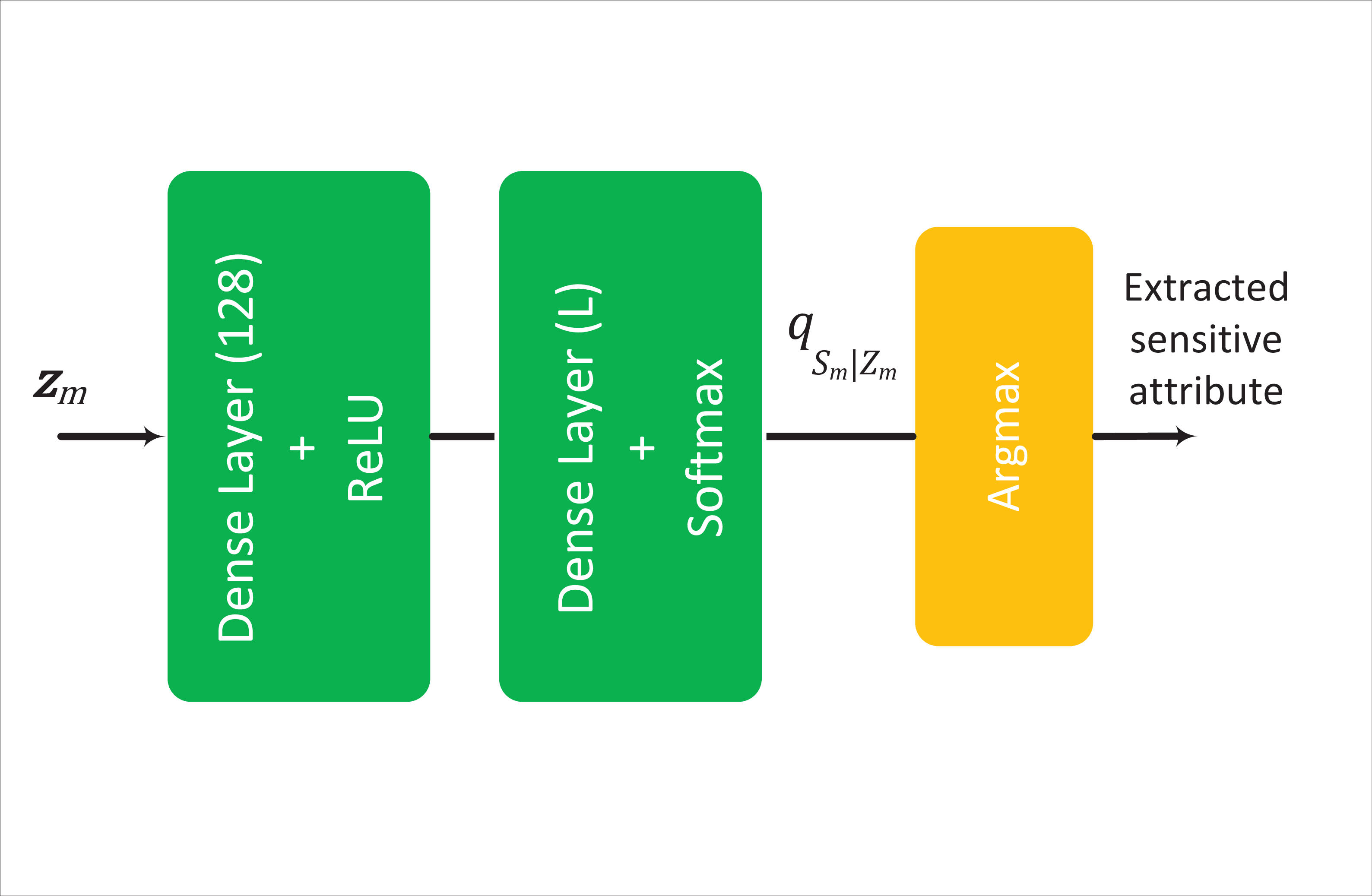}
		\vspace{-3mm}
		\caption{Implemented DNN at adversaries for extracting sensitive information.}
		\label{fig:DNN-Advs}
  \vspace{-4mm}
	\end{figure}

	\vspace{-0mm}
	According to the Deep-JSCC concept \cite{DJSCC-Deniz}, 
	we employ  autoencoder DNNs  to  directly map the  image pixels to  channel input symbols. 
	In  this regard, $\cal A$ maps  each realization of  the source data  $U^n$,  denoted by  $\bm  u \in \mathbb{R}^n$,   to a vector of   channel input   ${\bm x} \in \mathbb{C}^k$, which can be viewed as a realization of  $X^k$.  
	The block diagram of the DNNs employed for  the neural encoder and decoder components of  legitimate parties  is illustrated in Fig. \ref{fig:DNN}. 
	In addition, the DNN structure employed by each of the  adversaries is demonstrated in Fig. \ref{fig:DNN-Advs}. 
		
\vspace{0mm}

\vspace{-0.5mm}	
According to the proposed system model,  adversaries 
	utilize DNNs to facilitate the inference   of  sensitive  information $S$ from their received signals.  $S$ can be 
	the class labels of the images  \cite{AE-Deniz, Ecenaz-icassp}.    
	For example, the identity of patients within  medical imaging in e-health applications.    
	To 
	infer   sensitive attributes from images,  each adversary employs the DNN architecture illustrated in Fig. \ref{fig:DNN-Advs}, where the  dimension of the  output  neurons, $L$, equals the 
	cardinality  of the secrets $|S|$.\footnote{It would be interesting to 	study the  performance when the eavesdroppers employ more complicated neural  architectures, such as VGG and ResNet variation, to infer the secret. This will be addressed in our future works.} 
	The output of the softmax layer produces  an adversarial likelihood  estimation regarding the posterior  distribution  $q_{\Theta_{E,m}}({s_m|{\boldsymbol z}_m})$.  
	\vspace{-0.5mm}
	Invoking  \eqref{eq:P-DL},  one can say that each Eve  tries to minimize its CE between the adversarially-estimated posterior  distribution $q_{\Theta_{E,m}}({s_m|{\boldsymbol z}_m})$ and the ground-truth, which is represented by the one-hot encoded vector of $S$, denoted by  $ {{\bm \varepsilon}_s} \in \{0,1\}^L$.   
	Notably, having lower CE values results in  
	higher similarity between the  adversarial posterior distribution    	and the ground-truth, which increases the information leakage in terms of CE.    
	Meanwhile, $\cal A$ and $\cal B$  try to jointly  minimize the reconstruction distortion  and the information leakage  measured by the negative CE metric. 
Hence,  the  sample-wise  communication framework  can be reformulated as:   
	\begin{align}\label{eq:P-DL-CE}
		\text{minimize}	\hspace{2mm} & \mathcal{L}(\Omega_{\cal A},\Omega_{\cal B},\mathbf{\Theta}_E)  = 
		\mathbb{E}_{p(\bm{u},\bm{\hat{u}})} \left[	d\left(\bm u,
		\bm{\hat{u}}
		\right)
		\right] 
		\nonumber \\ 
		& \hspace{-2mm}+ 
		\frac{1}{M}	\hspace{-2mm} \sum_{m \in [M]} \hspace{-2mm}  w_m \hspace{1mm}
		\underset{\Theta_{E,m}}{\text{max}}
		\hspace{0.5mm}
		\left(- H\hspace{-1mm}\left(q_{\Theta_{E,m}}({s|{\boldsymbol z}_m}),{{\bm \varepsilon}_s}\right)\right), 
	\end{align}
	where $\hat{\bm u} = f_{\Omega_{\cal B}}(\bm y)$, and 
	$p(\bm u, \hat{\bm u})$ stands for the joint probability distribution of the original and the reconstructed image, taking into account the randomness in the input image and the channel. 
	Since  the true distribution $p(\bm u)$ is often unknown, we estimate the expected distortion measure  using samples ${\bm u}_j$ from an available  dataset ${\cal D}_u$ by computing $\mathbb{E}_{p(\bm{u},\bm{\hat{u}})} \left[	d\left(\bm u,f_{\Omega_{\cal B}}(\bm y)\right)
	\right] \approx \frac{1}{N_u}\sum_{\bm{u}\in \mathcal{D}_u} d(\bm u, \hat{\bm u})$, where  
	 ${N_{u}} \overset{\Delta}{=} |\mathcal{D}_u|$.
	{It is assumed  that we know the sensitive attribute in which the eavesdroppers are interested, as well as their channel models.  Both of these assumptions  are common  in the privacy \cite{AE-Deniz, Ecenaz-icassp} and wiretap channel \cite{Eduard-AE,  MI-Wiretap-estim, MI-estim-channel_coding} literature.  
	We do not need to know the instantaneous channel gains, but use their distributions to sample channel realizations during training.}     
	
	\vspace{0mm}
	\subsubsection*{Training Procedure}\label{sec:training}
	In order to train our  system  based on   \eqref{eq:P-DL-CE}, we  follow an  iterative procedure.  
	Intuitively, the network nodes  are faced  with a  minimax game, i.e., the competition between legitimate autoencoder and the adversarial  DNNs.  
	Hence,  the following strategy is run  through our proposed E2E system:  	
		The encoder and decoder function of Alice and Bob should jointly minimize their LF, denoted by $\mathcal{L}_{\cal AB}$:  
        \begin{align}
        & \hspace{0.5mm}\mathcal{L}_{\cal AB}  =
         \nonumber \\
         &  \hspace{0.5mm} \frac{1}{N_{u}}\hspace{-1.5mm}\sum_{{\bm u} \in \mathcal{D}_u}
			\hspace{-1.5mm} \left( \hspace{-1mm}
			d(\bm u, \hat{\bm u})\hspace{-1mm}-
			\hspace{-1mm}\frac{1}{M} \hspace{-2mm}\sum_{m \in [M]} \hspace{-3mm} w_m \hspace{0mm}
		H\hspace{-0.5mm}\left(\hspace{-0.5mm}q_{\Theta_{E,m}}\left({s{(\bm u)}|{\bm z}_m{(\bm u)}}\right),{{\bm \varepsilon}_{s}}{(\bm u)} \hspace{-0.5mm} \right)\hspace{-1.9mm}
			\right)\label{eq:L_AB}
        \end{align}
	The training process of legitimate nodes can be further enhanced via employing  \emph{adversarial likelihood compensation} (ALC), which has been shown in \cite{AE-Deniz} to be more effective in confusing an adversary than the one-hot encoding approach.   
		The main idea is to make the posterior distribution  of  adversaries 
		imitate a uniform distribution $\bar{p}_L = [\frac{1}{L},\cdots,\frac{1}{L}]^{\sf T}$. 
		Hence, {Alice and Bob jointly maximize the uncertainty of adversarial  predictions, resulting in 
		the following  loss function 
        \begin{align}
         & \hspace{0.5mm} \mathcal{L}^{\sf ALC}_{\cal AB} \hspace{-0.5mm}  = \nonumber \\
         & \hspace{0.5mm} \frac{1}{N_{u}}\hspace{-1mm}
		\sum_{{\bm u} \in \mathcal{D}_u} \hspace{-2mm}
		\left( \hspace{-1mm}
		d({\bm u}, {\hat{\bm u}}) \hspace{-1mm}
		+  \hspace{-1mm}
		\frac{1}{M} \hspace{-1.5mm} \sum_{m \in [M]} \hspace{-1.5mm} w_m
		H\hspace{-0.5mm}\left(q_{\Theta_{E,m}}({s{(\bm u)}|{\bm z}_m{(\bm u)}}),\bar{p}_L \hspace{-0.5mm}\right)\hspace{-1.0mm}
		\right) \label{eq:L_AB-LE}
        \end{align}
    	The distortion measure we consider for our legitimate loss function $	\mathcal{L}_{\cal AB}$ is a mixture of the average mean squared error (MSE), denoted by ${\Delta}^{\sf{MSE}}$, and 
    the structural similarity index (SSIM), ${\Delta}^{\sf{SSIM}}$,  between the  input image $\bm u$  and the recovered version $\hat{\bm u}$  at the output of Bob's DNN.
    Therefore, we assume $d(\cdot,\cdot)$ to be measured  as follows 
    \begin{align}\label{eq:distortion}
    	d(\bm u , \hat{\bm u}) = 
    	{\Delta}^{\sf{MSE}}(\bm u , \hat{\bm u}) + 
    	\alpha {\Delta}^{\sf{SSIM}}(\bm u , \hat{\bm u}),
    \end{align}
    where 
    $	\Delta^{\sf MSE} (\bm u , \hat{\bm u})
    \overset{\Delta}{=}\frac{1}{n}||\bm u-\hat{\bm u} ||^2$,  
    $\Delta^{\sf SSIM}(\bm u , \hat{\bm u}) \overset{\Delta}{=} 1- 	\mathsf{SSIM}(\bm u, \hat{\bm u}),  
    $  
    and $\alpha$ is a tuning parameter representing the contribution of the SSIM metric.  
    The SSIM measure between two images $I$ and  $K$ is defined as 
    \begin{align}\label{eq: ssim}
    	\mathsf{SSIM}(I, K) \overset{\Delta}{=} \frac{(2\mu_{I}\mu_{K}+c_1)
    		(2\sigma_{IK}+c_2)}{(\mu^2_{I}+\mu^2_{K}+c_1)(\sigma^2_{I}+\sigma^2_{K}+c_2)}, 
    \end{align} 
    where   $\mu_{I}$, $\mu_{K}$, $\sigma_{I}$, $\sigma_{K}$, and $\sigma_{IK}$ are the local means, standard deviations, and cross-covariance for images $I$ and  $K$, while $c_1$ and $c_2$ are two adjustable constants \cite{ssim-learning}.  
    The rationale behind 
    the  proposed distortion metric 
    is that we not only aim to recover every pixel of images with minimum error (captured via the MSE measure), but also want to obtain a 
    good-quality  reconstruction from the  \emph{human perception} point-of-view. 

		Each step of the  joint training of $\cal A$-to-$\cal B$ autoencoder  is  followed by a training step  for the adversarial DNNs.
		Eavesdroppers aim to minimize the CE between their estimated likelihood $q_{\Theta_{E,m}}({s|z_m})$ and the ground-truth vector ${{\bm \varepsilon}_s}$ corresponding to $S$. 
		Hence, the following  LF is employed  for   training  the DNN of  Eve$_{m}$ for $ m \in [M]$:     
		\begin{align}\label{eq:L_E}
			\hspace{-2mm}{\cal L}_{E,m} =  \frac{1}{N_{u}} 
				\hspace{-1mm}
			\sum_{{\bm u} \in \mathcal{D}_u}
				\hspace{-1.5mm}
			H\hspace{-1mm}\left(q_{\Theta_{E,m}}({s{(\bm u)}|{\bm z}_m{(\bm u)}}),{{\bm \varepsilon}_{s}}{(\bm u)}\right). 
		\end{align}  
	Note that the adversarial DNNs can be trained in parallel.  
	Then for the case of colluding eavesdroppers, an additional step of ``knowledge sharing'' is performed. In this case,  the adversaries share their individually-extracted logits, and a weighted sum of these logits is exploited for the inference of private attributes, where the logit  weights are  trained  in the colluding framework.

\begin{figure}
	\centering
	\includegraphics
	[width=3.0in,height=2.0in,
	trim={0.2in 0in 0 0.1in},clip]
	{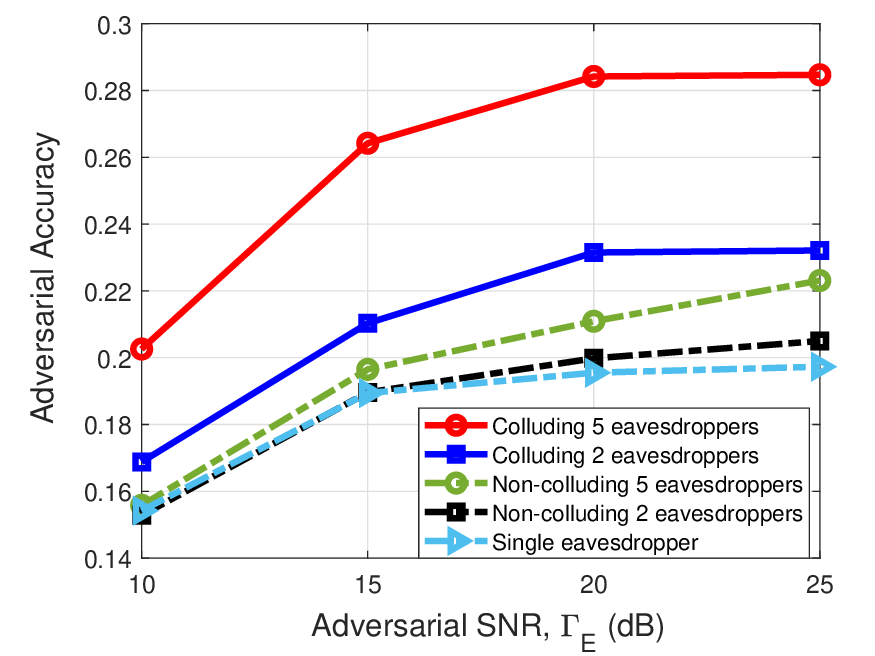} 
	\vspace{-1mm}
	\caption{{Total adversarial accuracy over Rayleigh channels.}}
	\label{fig:acc}
	\vspace{-3mm}
\end{figure}

\begin{figure}
		\centering
		\includegraphics
		[width=3.0in,height=2.0in, 
		trim={0.1in 0.0in 0 0.0in},clip]
		{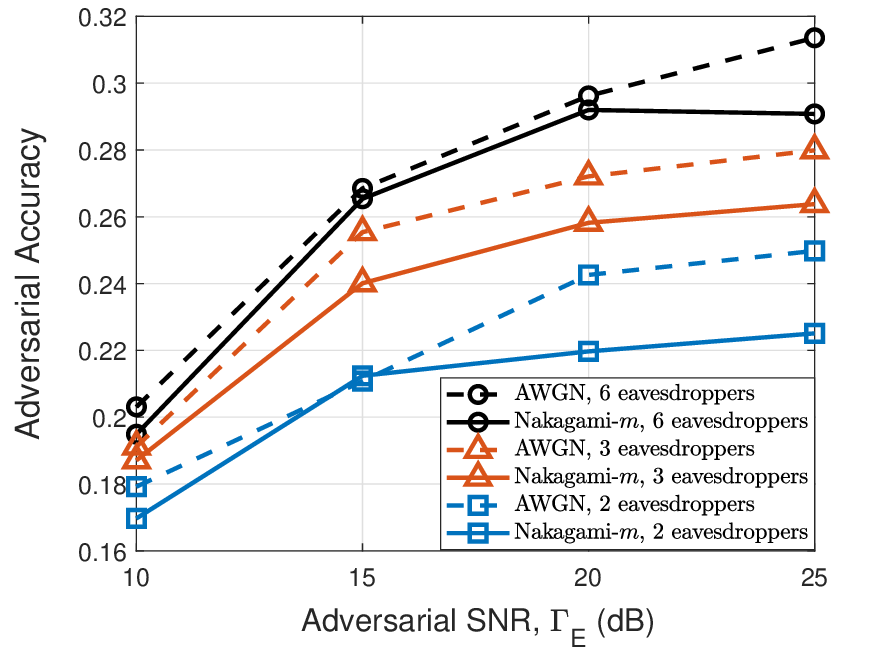}
		\vspace{-1mm}
		\caption{Colluding adversarial accuracy  over AWGN and Nakagami.}
		\label{fig:acc_Nak}
	\vspace{-5mm}
\end{figure}

	\vspace{0mm}
	\section{Evaluations}\label{sec:evaluation}
	\vspace{0mm}
	In this section, we evaluate  the performance of the proposed scheme over  both  AWGN and complex  fading (Rayleigh and Nakagami-$m$) communication channels.  
	We  address  the generalization capability of our proposed scheme for different communication scenarios and over a wide range of signal-to-noise ratio (SNR) values, to highlight the  \emph{data efficiency} of our proposed learning-based security  solution. 
	We also address the \emph{secrecy-utility trade-off} for the proposed learning-based approach. 
		We evaluate our proposed secure framework using images
		with dimension $32 \times 32 \times 3$ (height, width, channels) from CIFAR-10 dataset \cite{CIFARdataset}. 
 The dataset consists of 60000
		colored images of size $32 \times 32$ pixels. The training and
		evaluation sets are two completely separated sets of images,
		containing 50000 and 10000 images, respectively, associated
		with 10 classes.
	Adversaries wish to 
	infer a common secret $S$,  
	either individually (the non-colluding case), or by 
	learning from 
	the combination of the adversarial  logits
	(the colluding setup).
	The common secret $S$ here is considered as the  class of CIFAR-10 images with $|S| = L = 10$.  
	For simplicity, we consider a single weight in the LF, that is, $w_m = w = 5, \forall m$. We also set  $\alpha = 0.1$ and $n_{\sf T} = 4$ antennas. 
	These parameters are set after conducting extensive experiments and training the DNNs with a wide range of values for $w_m$ and $\alpha$, where we have omitted the  results of fine-tuning step due to space limitations.  
{Transmit SNRs of communication links are  defined as
	$	\Gamma_{B}=10\log_{10}\frac{P}{\sigma_{L}^2}$ dB and   $\Gamma_{E}=10\log_{10}\frac{P}{\sigma_{E}^2}$ dB, 
	 representing the ratio of the average power of the channel input to the average noise power of legitimate $\sigma^2_{L}$ and adversarial nodes $\sigma^2_{E}$, respectively. 
	During training,  we set $\Gamma_{B} = 20$dB and $\Gamma_{E} = 15$dB, respectively, while we test the performance over different values of channel SNRs during the inference.     
	 In addition, the bandwidth compression ratio 	is set to  $\frac{k}{n}=\frac{1}{3}$.    
	For the training, we sample channel realizations from the general case of complex Rayleigh  fading  model with average $\Gamma_{B}$ and $\Gamma_{E}$ values stated above.  	 
	Nevertheless, during the inference phase we study the performance  in  different scenarios of AWGN and Nakagami-$m$ channels (for $m=3$).  
 While we do assume  known channel models in our simulations, which we use to generate samples from conditional channel distribution, we could easily drop this assumption if we had data collected from a particular channel with unknown statistics.
	 DNN architectures are implemented  using  {Python3 with  Tensorflow.\footnote{https://www.tensorflow.org/}}  
{The codes  were run  
	on Intel(R) Xeon(R) Silver 4114  CPU running at 2.20 GHz with GeForce RTX 2080 Ti GPU.}
To minimize the LFs, 
 the widely-adopted Adam optimizer is chosen    \cite{adam} with a learning rate of $10^{-4}$. 
We  fix the number of training episodes to ${N}_{\text{episode}}  = 200$, and the batch size to $m=128$.

	Figs. 
\ref{fig:acc}
and
\ref{fig:acc_Nak}
show the adversarial accuracy of the  proposed scheme vs. the SNR of adversarial links ($\Gamma_E = \Gamma_{B} - 5 \text{ dB}$) for both scenarios of colluding and non-colluding  eavesdroppers. 
	For the colluding setup, the total adversarial accuracy refers to the overall accuracy of adversaries in correctly finding the ground-truth $\bm \varepsilon_s$ from their aggregated logits, while for the non-colluding benchmarks, the mean accuracy across eavesdroppers is plotted for the sake of comparison. 
	One can observe that increasing the number of eavesdroppers leads to   higher  accuracy for the adversaries, which is aligned with one's intuition.  
	The increase in adversarial accuracy is more significant in the colluding case due to the collaboration and ``knowledge sharing''  among eavesdroppers through the ensemble learning process \cite{ensemble}. This actually   helps them learn the secret more accurately. 
	The figures also indicate that by increasing the quality of adversarial links, i.e., increasing $\Gamma_{E}$, 
	the accuracy of adversaries  
	increases by 
	at most  $10\%$.  
	This is because  higher SNR values result in having less-distorted (less noisy) observations at the eavesdroppers, resulting  in  more accurate  estimations  about the  posterior adversarial distribution  $q_{\Theta_{E,m}}({s|{\bm z}_m})$. 
	The amount of increase in the adversarial accuracy reduces with the increase in $\Gamma_E$, which highlights the limitation of eavesdroppers in the  proposed secure scheme. 
	Fig. 
	\ref{fig:acc_Nak} 
	also highlights  the generalization capability of our learning-based  framework 
	extended to 
	the AWGN  and Nakagami-$m$ (for $m=3$) channel scenario. 
It shows that we can achieve almost similar  trends in channel scenarios other than Rayleigh fading, despite training the networks with a Rayleigh channel model. 

\begin{figure}
	\centering
	\includegraphics
	[width=3.0in,height=2.2in,
	trim={0in 0.0in 0 0.0in},clip]
	{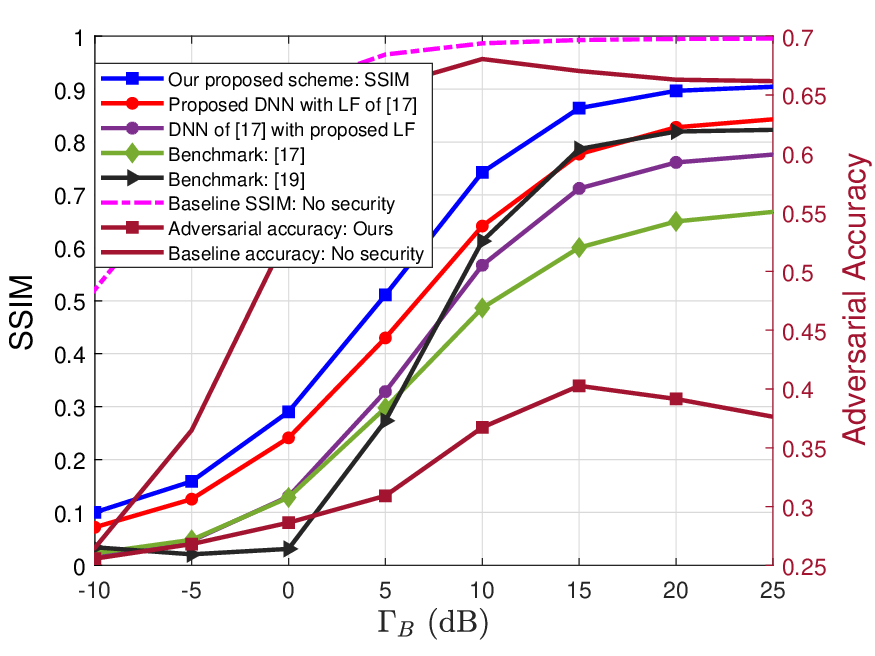}
	\vspace{-2mm}
	\caption{Ablation study for $M=3$ 
		non-colluding 
		eavesdroppers.}
	\label{fig: ssim_ablation_test}
	\vspace{-5mm}
\end{figure}

	\vspace{-0.5mm}
Fig. 
\ref{fig: ssim_ablation_test} 
illustrates the data   reconstruction performance at Bob and the total adversarial accuracy, when having $M=3$ non-colluding eavesdroppers.  
One can infer from the figure  that our proposed system outperforms the benchmarks in terms of the reconstruction performance. 
Accordingly, $20\%$ and $10\%$ performance gain is achieved by our proposed scheme 
compared with  \cite{AE-Deniz} and \cite{Deniz-GDN}, respectively.  
The ablation examinations conducted  in this figure show  that both the implemented DNNs   and the proposed LFs for optimizing the  framework contribute to the system's performance compared with other benchmarks.  
The figure also implies that increasing   $\Gamma_{B}$ results in having higher SSIM values. 
This is because increasing $\Gamma_{B}$ can result in   less distorted  observations at Bob,  which facilitates the image reconstruction performance.   
\emph{Data efficiency} and \emph{generalizability} of our proposed scheme are also validated, since we have trained our DNNs with a fixed SNR $\Gamma_{B} = 20$ dB, while the  performance gain  of our approach during  inference holds for various SNRs.  
Furthermore, Fig. 
\ref{fig: ssim_ablation_test} highlights that if we ignore the eavesdroppers during the training of $\cal A$-$\cal B$ pair, and set $w_m = 0$,  $\forall m\in [M]$, 
our proposed scheme can achieve almost perfect  (${\sf SSIM} = 1$) data recovery. The impact of the eavesdroppers on Bob's performance can be studied in this figure as well, where having $3$ eavesdroppers can impose $10\%$ decrease in the reconstruction performance of Bob. 
Finally, to indicate the importance  of our proposed adversary-aware scheme in terms of preventing leakage, we can observe from the figure that  if we do not employ secure neural encoding (i.e., ignoring the eavesdroppers during the training of $\cal A$-$\cal B$ pair), the adversarial accuracy is increased by about $28\%$. 
This clearly highlights the importance of employing our proposed learning-based secure encoding scheme.

\begin{figure}
	\centering
	\includegraphics
	[width=3.15in,height=2.0in,
	trim={0.0in 0.0in 0 0.2in},clip]
	{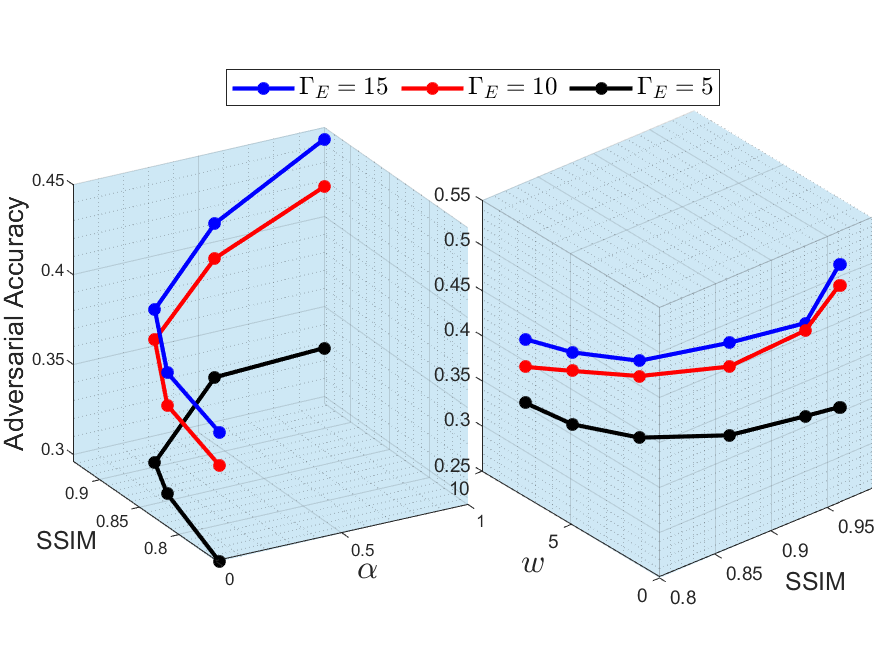} 
	\vspace{-4mm}
	\caption{Secrecy-utility trade-off.}
	\label{fig:3d_coeffs}
\vspace{-5mm}
\end{figure}

 Fig. \ref{fig:3d_coeffs} studies the impact of tuning parameters $\alpha$ and $w$  on the adversarial performance of our proposed system.  
These  hyper-parameters 
are
the coefficients associated with utility and secrecy   adjustment terms within our training LFs in 
\eqref{eq:distortion} and \eqref{eq:L_AB-LE}, respectively.   
For this experiment, the adversarial performance is captured by investigating the accuracy of adversaries in correctly finding the ground-truth label $\bm \varepsilon_s$ (representing the  sensitive information $S$) among the labels of  CIFAR-10 dataset.
The figure indicates that by increasing $\alpha$, higher values of SSIM can be achieved, since more emphasis on the SSIM error $\Delta^{\sf SSIM}$ is put based on \eqref{eq:distortion}. However, the accuracy of adversaries in extracting the sensitive information also increases, which verifies the secrecy-utility trade-off. 
Notably, by increasing  $\alpha$ to values  more than $0.1$, a jump in the adversarial accuracy   can be observed, which leads us choosing $\alpha = 0.1$ for our network.   
Similarly, by increasing $w$, the emphasis goes toward the secrecy criteria introduced  in   \eqref{eq:P0}, \eqref{eq:P-DL-CE}, and  \eqref{eq:L_AB-LE},  
which leads to the reduction in adversarial accuracy and achieved SSIM, verifying the secrecy-utility trade-off as well.  
Fig. \ref{fig:3d_coeffs} also  shows that increasing  $\Gamma_{E}$ can improve the  adversarial accuracy in finding the sensitive data 
$\bm \varepsilon_s$.  Notably, the  amount of increase in the adversarial accuracy reduces with the increase in $\Gamma_{E}$ which highlights the limitation of adversarial nodes based on our proposed  secure scheme.

	\vspace{-0.5mm}
	\section{Conclusions and Future Directions}\label{sec:Conclusion}
	\vspace{-1.0mm}  
	We proposed  a learning-based approach for E2E secure image delivery against multiple eavesdroppers  over AWGN and  complex-valued fading   channels.  
	{We considered both scenarios of {colluding} and {non-colluding} eavesdroppers
		over CIFAR-10 dataset.}  
	For the colluding  strategy, eavesdroppers collaborate  to infer private  data from their observations (channel outputs),  using ensemble learning, while for   the  {non-colluding} setup  they act alone.  
	Meanwhile, 	the legitimate parties aim to have a secure  communication with minimum average  distortion. 
	Employing  autoencoders, we proposed 
	a  secrecy funnel framework to  achieve  both secrecy and utility, where   we also take  into account the perceptual  quality of image transmission within our LF. 
	Evaluations  validate the performance  of our proposed scheme compared with existing benchmarks, while addressing the secrecy-utility trade-off.   
	Our proposed system is also shown to be \emph{generalizable} to a wide range of 
	SNRs and  different communication scenarios. 
	
	\emph{Future Directions:}  
	A problem to be studied is  
	training the systems over real-time wireless  channels. The challenge here  is the
	relatively long coherence time compared to the rate at which
	data samples can be processed for training.  Accordingly, 
	only a few channel realizations are observed  over every   minibatch,
	which will be an important issue for training communication systems that are supposed to generalize well to a wide range of channel realizations.

\end{document}